\documentclass[aip,jcp,floatfix,reprint]{revtex4-1}

\usepackage{graphicx}
\usepackage{dcolumn}
\usepackage{bm}
\usepackage{amsmath}
\usepackage{rotating}
\usepackage{amssymb}
\usepackage{verbatim}
\usepackage{float}
\usepackage{subfigure}
\usepackage{xcolor}

\begin{document}

\title{Necessary $N$-representability constraints from time-reversal symmetry for periodic systems}

\author{Nicholas C. Rubin}
\author{David A. Mazziotti}

\affiliation{Department of Chemistry and The James Franck Institute, The University of Chicago, Chicago, IL 60637 USA }

\date{Submitted April 8, 2016; Revised June 18, 2016}

\begin{abstract}
The variational calculation of the two-electron reduced density matrix (2-RDM) is extended to periodic molecular systems.  \textcolor{black}{If the 2-RDM theory is extended to the periodic case without consideration of time-reversal symmetry, however, it can yields energies that are significantly lower than the correct energies.  We derive and implement linear constraints that enforce time-reversal symmetry on the 2-RDM without destroying its computationally favorable block-diagonal structure from translational invariance.  Time-reversal symmetry is distinct from space-group  or spin (SU(2)) symmetries which can be expressed by unitary transformations.}  The time-reversal symmetry constraints are demonstrated through calculations of the metallic hydrogen chain and the one-dimensional lithium hydride crystal.
\end{abstract}

\maketitle

\section{Introduction}

Computation of the electronic structure of extended molecular systems is critical for elucidating the fundamental principles that govern the properties and behavior of molecules as they grow in size with applications throughout materials and biology.  Many large-scale molecules with favorable transport properties have energetically degenerate orbitals or spatial domains that give rise to strong electron correlation effects.  Strong electron correlation occurs when one or more electrons become entangled between two or more molecular orbitals.  Even without strong electron correlation the treatment of extended systems has computational challenges including the realistic incorporation of long-range Coulomb interactions.  The presence of strong electron correlation in the molecule or material places significant additional requirements on the complexity of the underlying quantum-mechanical wave function for an accurate solution, which limit the applicability of many traditional electronic structure methods.   Recently, accurate computations of strongly correlated molecules have been achieved through the direct calculation of the two electron reduced density matrix (2-RDM) rather than the many electron wave function~\cite{RDM07,CR12,coleman_reduced_Mat,RevModPhys.35.668, JMathPhys.8.2063, Erdahl,PhysRevA.63.042113, nakata_variational_2001, PhysRevA.65.062511,Zhao_T1,PhysRevLett.93.213001,nakata:164113,PhysRevA.80.032508,PhysRevLett.106.083001, PhysRevLett.108.263002}.  In the present paper we extend the variational 2-RDM theory to treat periodic molecular systems.

Because electrons are indistinguishable pairwise interactions, the ground or excited-state energy of an $N$-electron atom or molecule can be expressed as a linear functional of the 2-RDM~\cite{RDM07,CR12,coleman_reduced_Mat}.   In contrast to density functional theory variational 2-RDM theory has a known, linear functional.  However, direct minimization of the energy as a 2-RDM functional generates an energy that is significantly lower than the correct ground-state energy.  Additional constraints must be placed on the 2-RDM to ensure that it represents an ensemble $N$-electron quantum system.   These constraints are known as $N$-representability conditions~\cite{RDM07,CR12,coleman_reduced_Mat,RevModPhys.35.668, JMathPhys.8.2063, Erdahl,PhysRevA.63.042113, nakata_variational_2001, PhysRevA.65.062511,Zhao_T1,PhysRevLett.93.213001,nakata:164113, PhysRevA.80.032508, PhysRevLett.106.083001, PhysRevLett.108.263002}.  Recent work has developed a systematic hierarchy of ensemble $N$-representability conditions, containing previously known conditions, as well as the classical conditions~\cite{PhysRevLett.108.263002}. With an approximate set of these conditions the 2-RDM can be directly computed for strongly correlated quantum systems with polynomial-time computational complexity.  \textcolor{black}{The variational 2-RDM method has been applied to generating accurate energies and properties for} polyaromatic hydrocarbons~\cite{GM08,P11,DG16}, firefly luciferin~\cite{Greenman_firefly_2010}, transition-metal complexes~\cite{SHM16}, spin systems~\cite{hammond_variational_2005, anderson_second-order_2012, verstichel_variational_2012, baumgratz_lower_2012, barthel_solving_2012}, quantum dots~\cite{rothman_variational_2008}, and quantum phase transitions~\cite{gidofalvi_computation_2006,schwerdtfeger_convex-set_2009}.  \textcolor{black}{The most recent, largest calculations~\cite{PhysRevLett.106.083001,SHM16,DG16} are made possible from a two-orders-of-magnitude speedup from an efficient large-scale boundary-point algorithm~\cite{PhysRevLett.106.083001} for solving the semidefinite program~\cite{VB96, W97, NN93} associated with the variational 2-RDM optimization~\cite{Erdahl,PhysRevA.63.042113, nakata_variational_2001, PhysRevA.65.062511,Zhao_T1}.}

In this paper variational 2-RDM calculations are extended to periodic systems by building the periodicity into the definition of the molecular orbitals through crystalline-orbital Hartree-Fock calculations.\textcolor{black}{~\cite{JCP120.6.2581.Hirata, dovesi1988hartree}   Although crystalline-orbital Hartree-Fock theory has been previously extended to second-order many-body perturbation theory and coupled cluster theory,~\cite{Bartlett10.1063/1.471545, JCP120.6.2581.Hirata} it has not been previously applied to strongly correlated molecular systems.}   If the variational 2-RDM theory is extended to the periodic case without careful consideration of time-reversal symmetry, however, it can yields energies that are significantly below the correct energies.

\textcolor{black}{Time-reversal symmetry is not trivially satisfied by spin or symmetry adaptation in a momentum (complex-valued) basis set.  Because of the anti-unitary nature of the time-reversal operator, accounting for time-reversal invariance in a system differs from space-group symmetry
invariance or spin (SU(2)) invariance discussed in Refs.~\onlinecite{PhysRevA.72.052505,PhysRevA.72.032510}.  In a basis set with real-space function support the time-reversal symmetry operation exchanges the $\alpha$ ($\langle {\hat S}_{z} \rangle = +1/2$) and $\beta$  ($\langle {\hat S}_{z} \rangle = +1/2$) spins, leaving the spatial part of the orbitals unchanged~\cite{sakurai2014modern}.  It can be shown that spin adaptation in such a basis set is equivalent to generating basis functions with time-reversal symmetry.  In a momentum-space basis set, however, spin-adaptation does not imply time-reversal invariance because orbital linear momentum $+k$ is mapped to $-k$ by the anti-unitary complex conjugation part of the time-reversal operator. Without extra constraints relating Kramers pair ($k;-k$) orbitals (or observables) a variational solution of the 2-RDM will break time-reversal symmetry.}

\textcolor{black}{In contrast the spin and spatial symmetries in Refs.~\onlinecite{PhysRevA.72.052505,PhysRevA.72.032510}, which can be imposed by blocking the 2-RDM as well as the $^{2} Q$, $^{2} G$, $T_{1}$, and $T_{2}$ matrices, time-reversal symmetry cannot be imposed by basis-function adaptation without breaking the important block diagonal structure from translational symmetry~\cite{hammond_variational_2005, anderson_second-order_2012, verstichel_variational_2012, baumgratz_lower_2012,barthel_solving_2012}.  In this paper we preserve the computationally favorable block structure from translational symmetry by adding time-reversal symmetry as constraints to the semidefinite program~\cite{VB96, W97, NN93, PhysRevLett.93.213001, nakata:164113, PhysRevLett.106.083001}.  After development of the theory we demonstrate the important role of the time-reversal symmetry constraints in periodic calculations of one-dimensional hydrogen and lithium hydride crystals.}

\section{Variational $2$-RDM  Theory With Periodic Boundaries}

In variational $2$-RDM theory on extended systems the reduced Hamiltonian and density matrix are blocked according to the irreducible representations of the translational operator~\cite{PhysRevA.72.052505, cotton2008chemical}.   The reduced Hamiltonian is expressed as
\begin{align}\label{reducedH}
^{2}K_{ik_{i},jk_{j}}^{ak_{a},bk_{b}} =& \frac{1}{(K_{L}\times N)-1} \left( \delta_{ik_{i}}^{ak_{a}} h_{jk_{j}}^{bk_{b}}\hat{a}_{jk_{j}}^{\dagger}\hat{a}_{bk_{b}} \right. \nonumber \\
 +& \left.  \delta_{jk_{j}}^{bk_{b}}h_{ik_{k}}^{ak_{a}}\hat{a}_{ak_{a}}^{\dagger}\hat{a}_{ik_{i}} \right) \nonumber \\
+& V_{ik_{i},jk_{j}}^{ak_{a},bk_{b}}\hat{a}_{ak_{a}}^{\dagger}\hat{a}_{bk_{b}}^{\dagger}\hat{a}_{jk_{j}}\hat{a}_{ik_{i}}
\end{align}
where the indices are composite indices representing the band index and quasi-momentum index $k$, $K_L$ is the number of $k$-points sampled, $\hat{a}^{\dagger}$ ($\hat{a}$) are Fermionic creation (annihilation) operators, and $h_{jk_{j}}^{bk_{b}}$ and $V_{ik_{i},jk_{j}}^{ak_{a},bk_{b}}$ are the one- and two-electron integral tensors in the crystalline orbital basis.  The Hamiltonian is non-zero wherever $(k_{i} + k_{j} - k_{a}  - k_{b})\mathrm{mod}(2\pi) = 0$ is satisfied.  The metric matrices all have the same blocking structure as they must share the symmetries supported by the Hamiltonian.  In momentum space the structure of the $p$-positivity constraints remain the same as position space.  These constraints restrict the $(p+1)$ metric (or overlap) matrices of the form
\begin{equation}\label{metricOverlap}
M_{k} = \langle \psi | \hat{C_{k}}^{\dagger}\hat{C_{k}} | \psi \rangle
\end{equation}
to be positive semidefinite.  The operator $\hat{C_{k}}$, represents the set of \textit{$p$}-particle operators of momentum $k$ that form the $p$-particle basis functions from which the overlap matrix is obtained.  Considering rank-$2$ polynomials of creation annihilation operators in Eq.~[\ref{capoly}] and substituting into Eq.~[\ref{metricOverlap}] we generate a set of metric matrices which constrain the $k$-dependent probability distribution of finding two particles, two-holes, and a particle-hole pair to be positive semidefinite.
\begin{align}\label{capoly}
\hat{C}_{D_{k}} = \hat{a}_{j k_{a}}^{\dagger}\hat{a}_{i k_{b}}^{\dagger} \;\;\;\ (k_{a} + k_{b})\mathrm{mod}(2\pi) = k \nonumber \\
\hat{C}_{Q_{k}} = \hat{a}_{j k_{a}}\hat{a}_{i k_{b}}  \;\;\;\ (-k_{a} - k_{b})\mathrm{mod}(2\pi) = k  \nonumber\\
\hat{C}_{G_{k}} = \hat{a}_{j k_{a}}^{\dagger}\hat{a}_{i k_{b}} \;\;\;\ (k_{a} - k_{b})\mathrm{mod}(2\pi) = k
\end{align}
The indices in Eq.~[\ref{capoly}] are composite indices corresponding to a band and momentum index. The computational implementation of the variational energy minimization with respect to the $2$-RDM is formulated as a \textit{semidefinite program} (SDP). The program is constructed by considering the minimization of the linear energy functional, Eq.~[\ref{Elinear}]
\begin{align}\label{Elinear}
E = \mathrm{Tr}[^{2}K \cdot\; ^{2}D]
\end{align}
subject to the following constraints:
\begin{equation}\label{affineC}
A X = b \;\;\;\;X \succeq 0.
\end{equation}
$X \succeq 0 $ is the block representation of the reduced density matrices in Eq.~(\ref{blockRDM}) constrained to be positive semidefinite.
\begin{equation}\label{blockRDM}
X = \sum_{k}\oplus\;
\begin{pmatrix}
^{1}D_{k} & 0 & 0 &0 & 0\\
 0 & ^{1}Q_{k} & 0 & 0  & 0\\
0 & 0 &  ^{2}D_{k} &0 & 0\\
0 & 0 & 0 & ^{2}Q_{k} & 0\\
0 & 0 & 0 & 0 & ^{2}G_{k}
\end{pmatrix}
\end{equation}

The $A$ matrix in Eq.~(\ref{affineC}) contains the mapping relation generated by considering the image of the RDMs  in each $p$-particle metric space by rearranging the sequence of creation/annihilation operators subject to their anticommutation relations.
\section{Time-Reversal Equality Constraints} \label{sec:eq}
We augment the equality constraints on the $2$-RDM by considering the necessary equivalence of Kramers pair density matrix blocks.  These equalities are derived by considering the similarity transform of the one- and two-body operators with the time-reversal symmetry operator. The time-reversal symmetry operator is written as a product of a unitary operator and the complex conjugation operator with respect to a particular basis
\begin{align}
\Theta = UK
\end{align}
where the $U$ operator is the finite rotation around $y$-axis in spin-space by $\pi/2$ such that it anti-commutes with $\sigma_{x}$ and $\sigma_{z}$.
\begin{align}
U = \mathrm{exp}\left(-i\pi\;^{N}S_{y}\right)
\end{align}
Considering the similarity transform ($\Theta \hat{O} \Theta^{-1} = \hat{O}$) of a general one- and two-body operator with $\Theta$ we derive the equality constraints.  For the $1$-particle equalities we explicitly include the following relation in the $A$ matrix
\begin{align}
\Theta(^{1}D^{k}_{ij})\Theta^{-1} = \;[^{1}D_{ij}^{-k}]^{*}
\end{align}
\begin{align}
\Theta(^{1}Q^{k}_{ij})\Theta^{-1} = \;[^{1}Q_{ij}^{-k}]^{*}
\end{align}
where we have dropped the spin variable because spin-restriction requires $\alpha = \beta$. Unlike in position space where time-reversal symmetry forces the one-body operator to be real-valued~\cite{fukutome1981unrestricted} in a spin-adapted basis set, a general complex momentum space one-body operator can be complex valued as long as the matrix blocks corresponding to $(k,-k)$ pairs are complex conjugates of each other. For two-body density matrices we generate the constraints by applying time-reversal to each $\hat{C}$ operator
\begin{align}
\langle\psi|\Theta \hat{C}_{ij} \Theta^{-1} \Theta \hat{C}_{ab}^{\dagger} \Theta^{-1} |\psi\rangle
\end{align}
which results in the following equalities for each $(k,-k)$ block in the $2$-particle metric matrices
\begin{align}
^{2}M_{i\alpha,j\beta;k\alpha,l\beta}(k) = \left[ \;^{2}M_{i\beta,j\alpha;,k\beta,l\alpha}(-k) \right]^{*}
\end{align}
where $M = D,Q,G$.
When using spin-adapted $\hat{C}$ operators we arrive a well known expression mapping $(k,-k)$ blocks to each other for the singlet and triplet blocks~\cite{sakurai2014modern,PhysRevA.72.052505}.  A consequence of the above symmetries is that the $k = 0$ and $k = \pi$ blocks must be real valued on the one- and two-particle space.  Using the normal $p$-positive set, commonly denoted DQG, and including the time-reversal constraints we generate the approximate $N$-representability constraints used in this work.  All calculations using this set are labeled RDM-TR.  We compare the augmented $N$-representability constraints against normal $2$-positivity without the time-reversal equalities which are labeled RDM~\cite{Gidofalvi3pos}.

In the SDP the complex valued $k$-space density matrices are represented by $2N\times2N$ real matrices where $N$ is the linear dimension of the $k$-space matrix~\cite{ComplexValAsMatRef}.  The time-reversal (TR) equality constraints manifest themselves differently for symmetric or antisymmetric matrices.  The TR operator maps a geminal to its time-reversed pair $(ik_{i}, jk_{j}) \rightarrow (i\overline{k_{i}}, j\overline{k_{j}})$, where $\overline{k_{i}} = -k_{i}$, and thus occasionally we must employ the antisymmetry property as the matrix index of $i\overline{k_{i}}$ may be larger than $j\overline{k_{j}}$ in this case we have the RDM elements equal and opposite.  For these cases the imaginary component for the two density matrix elements is necessarily zero.
\section{Applications} \label{sec:app}
We test these constraints by calculating the binding energies of two one-dimensional polymers.  Each polymer is first described at the mean-field level with crystalline orbital Hartree-Fock (CO-HF)~\cite{JCP120.6.2581.Hirata, dovesi1988hartree}.  CO-HF performs Hartree-Fock on a set of non-orthogonal Bloch vectors built by Fourier summation of atomic orbitals over a super-cell.  The complex-valued crystalline orbitals are then used to build the one- and two-electron integral tensors in $k$-space~\cite{Bartlett10.1063/1.471545} which are subsequently used to build the reduced-Hamiltonian.  The CO-HF calculation performs a lattice truncation asymmetrically resulting in the destruction of the four-fold two-electron integral symmetry~\cite{szabo2012modern, JCP120.6.2581.Hirata}.  This symmetry along with the time-reversal symmetry for each $k$-point is explicitly restored when building the reduced Hamiltonian prior to the calculation.  This is accomplished by explicitly building the two-electron integral tensor with eight-fold symmetry--four from the integrals times two from time-reversal.

For variational minimization of the energy with respect to the $2$-RDM subject to $2$-positivity conditions that are not augmented with the time-reversal symmetry constraints we find the calculations are either i) not able to find a ground state solution or ii) converge to a $2$-RDM with broken TR symmetry.  This results in an energy that is below the true ground state energy.

The first system we consider is the binding of an infinite hydrogen chain.  We compare the energies from CO-HF, MP2, variational RDM with DQG constraints and variational RDM with DQG plus time-reversal equality constraints around the chains binding minimum.  For the crystalline orbital Hartree-Fock we use a unit cell of two hydrogen atoms and set the short and long range cut off criteria to be 10.  We sample $k$-space at 20 evenly spaced points.  M{\o}ller-Plesset perturbation theory for the periodic system is implemented based on Refs.~\onlinecite{Bartlett10.1063/1.471545,JCP120.6.2581.Hirata}.
\begin{figure}
    \includegraphics[width=8.5cm]{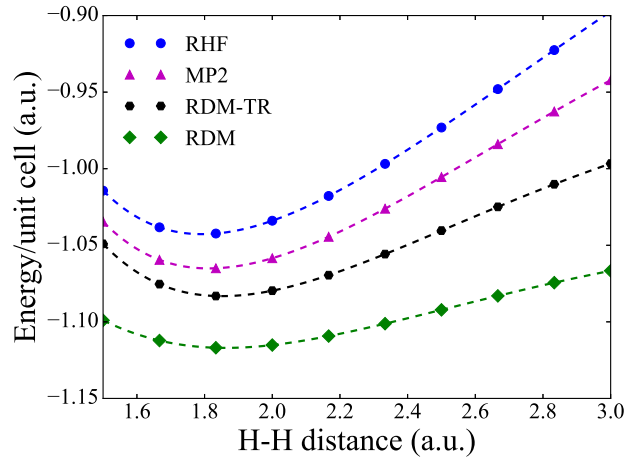}
    \caption{(H$_{2}$)$_{\infty}$ computed at RHF, MP2, RDM, and RDM-TR.  \textcolor{black}{When the additional time-reversal symmetry constraints are added to the SDP we recover a solution that corresponds with large-scale open boundary condition calculations for the entire binding region.  The RDM-TR curve is on average within a few mhartrees of a 50-atom DMRG calculation.~\cite{Chan.jcp.125.14.10.1063}  \label{H2_bind}}}
\end{figure}
\begin{figure}
    \includegraphics[width=8.5cm]{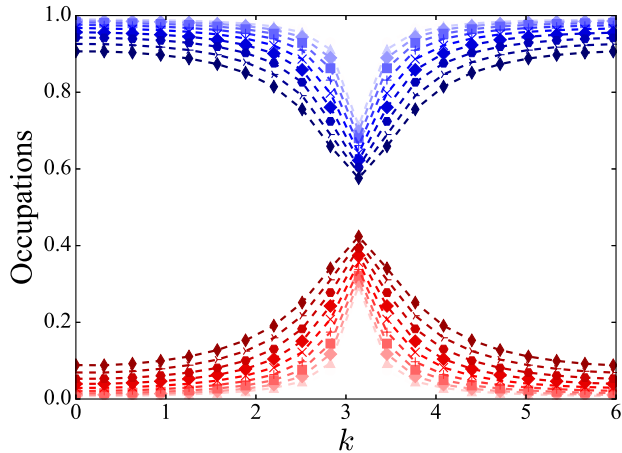}
    \caption{Occupation numbers of RDM with time-reversal symmetry.  Symmetric around $\pi$ and approaching the correct  completely dilated lattice values of 0.5.  Blue curves are the HONO occupations and red curves are LUNO occupations.  The darker the curve indicates a more dilated lattice. \label{H2_bind_occs}}
\end{figure}
It has been previously demonstrated that correlations in hydrogen chains cause the density matrix elements to decay extremely slowly with respect to the central unit cell~\cite{JCPYamadaHirataAsymptotic2013}.  Despite this, their energies and properties, such as conduction, converge extremely rapidly with respect to chain size~\cite{QUA:QUA23047}.  Therefore, we can evaluate the efficacy of the $k$-space variational $2$-RDM method by evaluating the accuracy of the hydrogen chain at the dissociated limit and the binding minimum by comparing the energy of PBC calculations against the energy for analogous large-scale finite chains with open boundary condition.  We plot the energies in Fig.~[\ref{H2_bind}] for the four aforementioned methods.  MP2 shows good improvement over the Hartree-Fock solution but is known to diverge as the system becomes more correlated.  The energy determined by variational $2$-RDM theory with normal $2$-positivity constraints results in a lower bound of approximately 50 milliHartrees for the entire binding region of the hydrogen chain.  When the additional time-reversal symmetry constraints are added to the SDP we recover a solution that corresponds with large-scale open boundary condition calculations for the entire binding region.

\begin{figure}
\centering
\includegraphics[width=8.5cm]{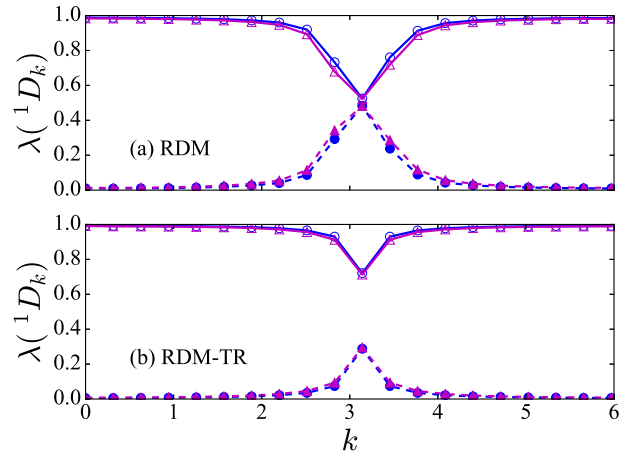}
\caption{Metallic solution $k$-dependent occupation numbers at points 0 and point 1 (1.5 bohr and 1.666 bohr separation of hydrogen atoms in the chain) on the binding scan above at the RDM level.  The top (a) is no time-reversal symmetry.  The bottom (b) is time-reversal symmetry restored.  The restoration of the symmetry enforces the correct density matrix symmetry around $\pi$.\label{MetallicKH2}}
\end{figure}

Time-reversal symmetry dictates a degeneracy in the density matrices and thus eigenvalues at the $1$- and $2$-particle level\cite{PhysRev.147.896, FreedNrep}.  In Fig.~[\ref{MetallicKH2}] we compare the occupation numbers for each band at each $k$-point for two metallic solutions.  The variational RDM calculation that is un-augmented with TR constraints produces occupations that are not only asymmetric around $\pi$ but also artificially large.  When TR equality constraints are added the occupations are symmetric around $\pi$ as expected.  Fig.~[\ref{H2_bind_occs}] is a plot of all the $k$-point occupations for the dilation of the hydrogen chain.  We see that as we approach the dilated limit the occupations approach the physically correct value of 0.5.

In many cases it is unnecessary to correlate core electrons.  For the binding curves of the lithium hydride chain we consider an active space of bands around the Fermi surface.  The active space Hamiltonian treats core electrons at the mean-field level without relying on pseudo-potentials for the valence~\cite{Shavitt1.430426}.  We selected all bands involving significant character on the frontier orbitals of the unit cell.  With just three bands in the active space we are able to capture the correct dissociation character of the lithium hydride crystal.  The lithium hydride crystal was built by considering a single lithium hydride in the unit cell with five neighboring cells.  The Li-H distance was $4.0$ Bohr while the H-Li distance was $6.0$ Bohr.
\begin{figure}
\includegraphics[width=8.5cm]{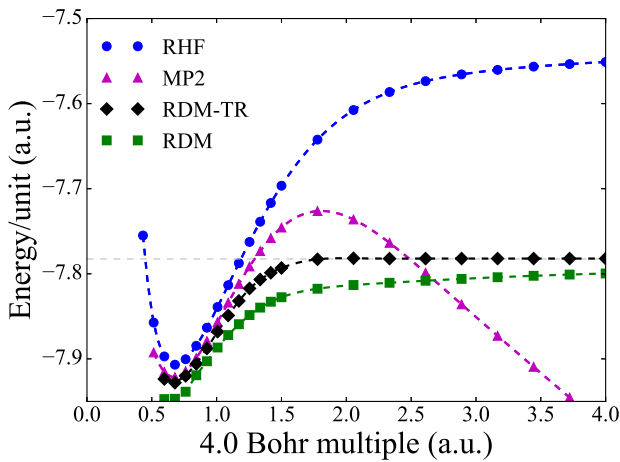}
    \caption{(LiH)$_{\infty}$ computed with RHF, MP2, and RDM .  The RDM calculation with and without time-reversal symmetry $N$-representability three active bands. The grey line is a (LiH)$_{6}$ dilated to its dissociated limit.  Despite CO-HF converging at lattice multiples less than 0.6 the CO-HF basis is so symmetry broken a good reduced Hamiltonian could not be generated.  As such, the RDM calculations without TR constraints failed to converge in this region.   \label{LiH_bind}}
\end{figure}
The crystal dilation was performed by multiplying the internal bond lengths by a scalar referenced in Fig.~[\ref{LiH_bind}].  Even on a small active space we see that time-reversal constraints are necessary to accurately describing the binding of LiH. Without the TR-equality constraints the RDM solutions lower bounds the correct solution throughout the binding region by a significant amount.

\section{Discussion and Conclusions}

Direct variational calculation of the 2-RDM, which is applicable to strongly correlated systems with polynomial computational scaling, was extended to periodic systems.   Additional constraints beyond space-group symmetry invariance or spin (SU(2)) invariance, we showed, are necessary to prevent the breaking of time-reversal symmetry.  Unlike the case for spin symmetry, basis-function adaptation for time-reversal symmetry destroys the computationally favorable block-diagonal structure of the 2-RDM from translational invariance.  In lieu of basis-function adaptation, in section~\ref{sec:eq} the necessary linear constraints on the 2-RDM were derived and incorporated into the variational calculation of 2-RDM by large-scale semidefinite programming.  In section~\ref{sec:app} we demonstrated the important role of the time-reversal symmetry constraints in periodic calculations of one-dimensional hydrogen and lithium hydride crystals where they restored the correct dissociation limits.

\textcolor{black}{The importance of symmetry breaking and restoration for describing strong (multi-reference) correlation has recently been recognized in several different contexts.  Scuseria and co-workers,~\cite{S12} for example, have employed symmetry breaking and restoration to recover multi-reference correlation effects at mean-field-like computational cost.  Veeraraghavan and Mazziotti~\cite{VM14a,VM14b} have examined the association of strong electron correlation with multiple, symmetry-broken  Hartree-Fock solutions, obtained from global solutions at different molecular geometries.  They suggested the use of these multiple Hartree-Fock solutions in a non-orthogonal configuration interaction (NOCI).  Burton and Thom~\cite{BT16} have recently performing NOCI on multiple symmetry-broken Hartree-Fock solutions, showing accurate agreement with full configuration interaction.  In the calculations presented here, the breaking of time-reversal symmetry generates additional distinct solutions, which in the position-space representation extend nontrivially from the real-axis into the complex plane.  Even though the focus of the present paper is to restore the time-reversal symmetry, the time-reversal symmetry-broken solutions may also contain useful, additional information about the electronic structure of the system.}

Accurate calculations of the correlated ground-state electronic structure of periodic and extended systems, especially in the presence of strong electron correlation, are significant for understanding and quantifying correlation-driven phenomena.  \textcolor{black}{The variational 2-RDM theory has been successfully applied to treat the strong electron correlation in quantum systems that are too large for conventional treatments including applications in both chemistry and physics.~\cite{GM08,P11,DG16,Greenman_firefly_2010,SHM16,hammond_variational_2005, anderson_second-order_2012, verstichel_variational_2012, baumgratz_lower_2012,barthel_solving_2012,rothman_variational_2008,gidofalvi_computation_2006, schwerdtfeger_convex-set_2009}  In this paper a significant theoretical bottleneck in treatment of periodic molecular systems is resolved through the derivation and inclusion of constraints on the 2-RDM to preserve time-reversal symmetry, which will make possible the application of 2-RDM theory to strongly correlated molecular systems with periodic symmetry.  Because the crystalline-orbital basis contains resolution in terms of both Gaussian orbitals and Fourier modes, the method can be applied to computing both local and long-range properties.}  The development of time-reversal constraints for 2-RDM calculations represents an important step in the direction of developing more accurate density-matrix-based electronic structure methods for \textcolor{black}{strongly correlated} periodic and extended systems with important potential applications in chemistry, physics, and materials science.

\begin{acknowledgments}

D.A.M. gratefully acknowledges the U.S. National Science Foundation CHE-1565638 the U.S. Army Research Office (ARO)  Grant No. W911NF-16-1-0152 and W911NF-16-C-0030, and the U.S. Air Force Office of Scientific Research (AFOSR) FA9550-14-1-0367 for their support.

\end{acknowledgments}

\bibliography{TRRDMR1}

\end{document}